**Scintillator-integrated microchannel plate photomultiplier tubes for ultrafast timing over keV–GeV energy scales**

Ryosuke Ota, [1,2,*] Yuya Onishi,[2] Daehee Lee, [1] Yuki Ichikawa, [3] Koji Kuramoto, [3] Kenshi Shimano, [3] Yutaka Hasegawa,[3] Eric Berg, [1] Takahiro Moriya,[2] Simon R. Cherry,[1] and Sun Il Kwon[1,†]

[1]Department of Biomedical Engineering, University of California, Davis, CA 95616, USA
[2]Central Research Laboratory, Hamamatsu Photonics K.K., Hamamatsu 434-8601, Japan
[3]Electron Tube Division, Hamamatsu Photonics K.K., Iwata 438-0193, Japan
*Contact author: ryoota@ucdavis.edu, ryosuke.ota@crl.hpk.co.jp
†Contact author: sunkwon@ucdavis.edu

**ABSTRACT**. Precise measurement of radiation has long played a vital role in a wide range of research and industrial fields, from fundamental physics beyond the Standard Model to medical imaging such as time-of-flight positron emission tomography. Developing radiation detectors that achieve high timing precision−on the order of a few tens of picoseconds–and energy measurement capabilities remains indispensable yet challenging. In this study, we developed two types of scintillator-integrated microchannel plate photomultiplier tubes (SCI-IMPs), one incorporating barium fluoride, and the other bismuth germanate, to enable simultaneous high-precision timing and energy measurements. To evaluate their performance over a wide energy range from keV– to GeV–scale, electron-positron annihilation gamma rays and cosmic ray muons were used. For energy measurements, both detectors achieved an energy resolution of approximately 35% at 511 keV. For timing measurements using 511 keV, coincidence time resolutions (CTRs) of approximately 50 ps full width at half maximum (FWHM) were obtained for both detectors. In contrast, for cosmic ray muon experiments where cosmic ray muon energy is typically on the order of GeV, CTRs were measured to be 25.1 and 16.8 ps FWHM for barium fluoride- and bismuth germanate-based detectors, respectively. The versatile scintillator-integration technique established in this study can broaden the applicability of the newly developed SCI-IMPs. In particular, these results demonstrate that the developed detectors push the boundaries of timing performance while retaining energy measurement and hold promise for future applications in fundamental physics experiments and medical imaging.

## I. INTRODUCTION

Radiation detection is one of the most indispensable techniques for human life in terms of exploring new science, which makes our lives safer and more sustainable. In the search for new physics, accelerator-based experiments are commonly utilized to artificially generate extreme conditions, and subsequent radiation from collisions and interactions between energetic particles at the GeV scale and above is detected and analyzed to elucidate how particles interacted with each other and to discover new physics beyond the Standard Model. When analyzing data, each subsequent particle produced after interactions is preferably temporally distinguished and separated depending on the origin of the interaction, in order to reconstruct particle tracks. A targeted timing performance is on the order of a few tens of picoseconds (ps) [1-5]. In addition, particle identification performance can also benefit from fast-timing information [6,7]. In this way, ultrafast timing information is essential for high-energy physics experiments.

Separately, in the medical field, time-of-flight positron emission tomography (TOF-PET) is being increasingly utilized in hospitals due to its quantitative diagnostic capability in life-threatening diseases such as cancer and Alzheimer's disease [8]. TOF-PET also makes full use of radiation detectors with high timing performance to estimate the locations of positron annihilations occurring in a patient's body by detecting pairs of 511 keV gamma-rays, thereby enhancing the signal-to-noise ratio of PET images and improving diagnostic accuracy, lesion detectability, and staging [8,9]. Currently, commercially available TOF-PET scanners have system-level coincidence time resolution (CTR) as fast as 200 ps full width at half maximum (FWHM) [10]. Therefore, a reasonable next goal is to achieve CTR < 100 ps FWHM [10-12] to enhance the information obtained from radiotracers. Developing detectors with ultrafast timing is needed to meet this goal.

Radiation detectors commonly consist of two main components that enable them to simultaneously detect energy and timing information: scintillators, which indirectly convert impinging particles into visible photons, and photodetectors, which detect these visible photons and convert them into electrical signals. Therefore, the timing performance of a radiation detector is largely determined by the physical properties of these components, such as the decay and rise time constants and light yield of scintillators, as well as the quantum efficiency (QE) and single photon time resolution (SPTR) of photodetectors [13]. In addition, adequately designing the dimensions of the scintillator is also important, because the photon transit time spread within the scintillator is on the order of tens of ps if the scintillator size is less than $10^3$ mm$^3$, which also limits the timing performance of detectors [14]. Therefore, achieving timing at the level of a few tens of ps is challenging, even when the intrinsic temporal properties of the scintillator and photodetector are better than the target time resolution. For more efficient photon propagation, the coupling method between a scintillator and a photodetector also plays an essential role. As an example, although barium fluoride (BaF$_2$) is an old but well-known scintillator for fast timing [15], and its intrinsic CTR should be better than 10 ps FWHM [16], its timing capability has not been fully exploited due to the difficulty of extracting and detecting ultraviolet photons around 220 nm.

Recently, in the PET field, detecting Cherenkov photons produced in a scintillator has received attention as an alternative method to extract precise timing information while maintaining energy detectability, because the physical mechanism of Cherenkov emission is faster than that of scintillation emission [17-19]. Bismuth germanate (BGO) is a long-established and inexpensive scintillator; moreover, it is recognized as a hybrid Cherenkov/scintillation material. Although numerous studies have aimed to maximize the potential of BGO by detecting its Cherenkov photons, it remains challenging for two main reasons: *(i)* ultraviolet Cherenkov photons are emitted more frequently than visible ones, and *(ii)* the number of Cherenkov photons emitted is approximately 270 times lower than that of scintillation photons at 511 keV. Therefore, the efficient propagation of Cherenkov photons toward the photodetector, prior to the arrival of scintillation photons, is crucial to achieving the target timing resolution.

*Contact author: ryoota@ucdavis.edu, ryosuke.ota@crl.hpk.co.jp

†Contact author: sunkwon@ucdavis.edu

A potential technological approach that can overcome the aforementioned difficulties is to integrate the Cherenkov/scintillation material directly as part of the photodetector. By integrating the Cherenkov/scintillation material directly into the window faceplate (WFP) of a photomultiplier tube (PMT), the optical boundaries between the material and the photocathode of a PMT can be eliminated. This configuration increases photon propagation efficiency, improves the detection probability of ultraviolet photons, and thereby enhances timing performance, as it prevents optical photons from being reflected back into the material. Cherenkov-radiator-integrated microchannel plate PMTs (CRI-MCP-PMTs) have demonstrated enhanced timing performance, achieving a CTR of approximately 30 ps using pairs of 511 keV gamma-rays [20]. However, these studies focused solely on the detection of Cherenkov photons, and the necessary energy information was compromised. Therefore, the applicability of such detectors is very limited, although MCP-PMTs in general have attracted attention for use in high energy physics and PET [21].

In this study, we optimized the integrating technique of scintillators with MCP-PMTs to retain all necessary information while providing excellent timing performance on the order of a few tens of ps. We developed two types of scintillator-integrated MCP-PMTs (SCI-IMPs): one with $BaF_2$ and the other with BGO. Their CTRs were experimentally evaluated using not only keV–scale gamma rays but also GeV–scale cosmic-ray muons. CTRs on the order of tens of ps FWHM, applicable across a wide range of energy scales and applications, are presented in the following sections.

## II. METHODS

This section first explains the concept of the newly developed SCI-IMPs, followed by details of the experimental setups from the keV– to GeV–scale.

### A. Integration technique of scintillators

An MCP-PMT was selected as the photodetector into which the scintillators were integrated because it is one of the fastest vacuum-based photodetectors with single photon detection capability, and its SPTR can be as fast as 25 ps FWHM. The WFP of the MCP-PMT is typically a transparent material such as synthetic glass or magnesium fluoride, and the photocathode is deposited on the vacuum side of the WFP. In a previous study [20], the WFP was replaced with lead glass as a pure Cherenkov radiator to make the MCP-PMT sensitive to gamma rays and to eliminate the optical boundaries between the lead glass and the photocathode. However, direct deposition of the photocathode onto the lead glass caused chemical reactions, turning the radiator black and making it insensitive to photons. To prevent these chemical reactions, we introduced an $Al_2O_3$ intermediate layer with a thickness of a few nanometers between the lead glass and the photocathode, thereby recovering sensitivity to photons. However, the previous study focused solely on Cherenkov photon detection and lacked energy information from gamma rays, which is indispensable in many applications, including TOF PET.

In this study, we developed two types of SCI-IMPs: $BaF_2$-IMPs and BGO-IMPs, in which the WFP was replaced respectively with $BaF_2$ or BGO, using the same technology as the lead glass Cherenkov radiator, to acquire energy information while maintaining fast timing capability. The dimensions of the $BaF_2$ and BGO WFPs are approximately 22 mm$\phi$ × 3.2 mm thick, however, the photocathode was deposited only on the central 11 mm$\phi$ region [20]. The basic characteristics of the PMTs and the scintillation properties of the developed detectors are summarized in Table I and visualized in Fig. 1.

An appropriate choice of photocathode, depending on the integrated scintillator, can maximize detector performance. The emission spectrum of the $BaF_2$ scintillator can be approximately divided into two components: a fast decay component with a shorter wavelength (~220 nm) and a slow decay component with a longer wavelength (~300 nm). Although the slow component has a higher light yield than the fast one, its presence not only degrades timing performance but also causes pile-up between temporally adjacent events, making the baseline unstable. To suppress this, a cesium telluride (CsTe) photocathode, which is sensitive to photons with wavelengths shorter than 300 nm, was employed. The spectroscopic QE curve of the CsTe photocathode is shown in the top of Fig. 1. The two QE curves differ due to fabrication variability, but both are approximately 17% at the emission peak (220 nm) of the fast component.

*Contact author: ryoota@ucdavis.edu, ryosuke.ota@crl.hpk.co.jp

†Contact author: sunkwon@ucdavis.edu

In the case of the BGO-IMP, a multialkali photocathode was selected based on the BGO emission spectrum, which is broad and has a cut-off wavelength of approximately 300 nm. The bottom plot in Fig. 1 shows the QE curves of two BGO-IMPs. The rapid drop in QE around 300 nm corresponds to the optical cut-off wavelength of BGO. The difference in the absolute values of the QEs is attributed to fabrication variability. The BGO-IMPs with high and low QE are denoted as A and B, respectively (e.g., BGO-IMP A).

The SPTR of MCP-PMTs strongly depends on the configuration of the divider circuit. Therefore, we also optimized the divider circuit in accordance with a previous study [22]. Specifically, the bias voltages between the photocathode and $MCP_{in}$ and between $MCP_{out}$ and the anode were set to approximately 550 and 1100 V, respectively.

TABLE I. PMT and scintillation properties of the developed SCI-IMPs.

|  | BaF$_2$-IMP | BGO-IMP |
|---|---|---|
| WFP | Barium fluoride | Bismuth germanate |
| Photocathode | CsTe | Multialkali |
| Max QE (%) | 19.6% at 240 nm | 20.9% at 400 nm |
| Density (g/cm$^3$) | 4.89 | 7.13 |
| Decay time constants (ns)[a] | 0.078 (0.99%) 0.747 (5.35%) 689 (93.66%) | 45.8 (8.2%) 365 (91.8%) |
| Light yield (photons/keV)[a] | 8.5 | 10.7 |

[a]Reference [15].

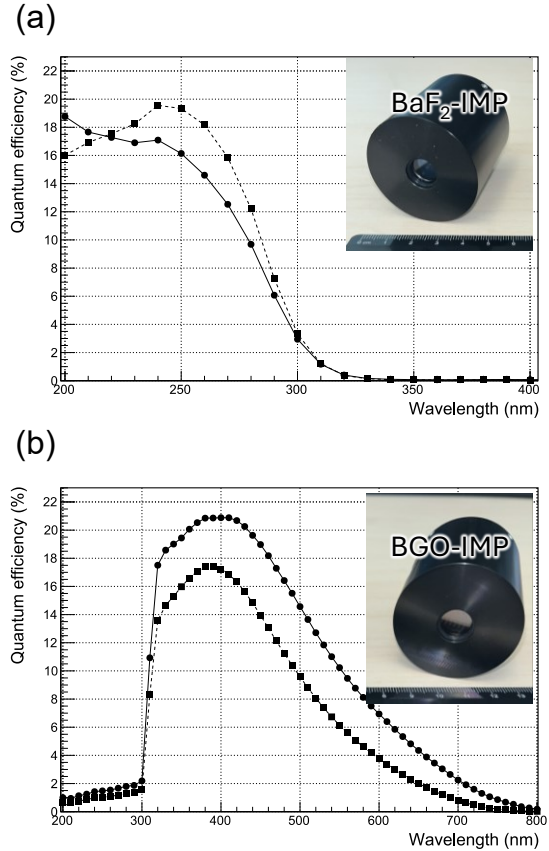

FIG. 1. Photograph of the newly developed SCI-IMPs with (a) BaF$_2$ and (b) BGO integrated as WFPs. The QEs of each SCI-IMP are also plotted as a function of wavelength., with two QE curves for each SCI-IMP representing the different properties developed in this study. For the BaF$_2$-IMPs, a CsTe photocathode was deposited to suppress the slow scintillation emission component of BaF$_2$, whereas a multialkali photocathode was used for the BGO-IMPs to provide sensitivity over a wide range of photons, including both Cherenkov and scintillation photons.

**B. CTR measurement using 511 keV gamma-rays**

Figures 2(a, b) display the experimental setups for measuring the CTR based on 511 keV gamma-rays using a $^{22}$Na point source. As a common setup for the two different experiments, tungsten collimators with an aperture size of $2 \times 2$ mm$^2$ and a thickness of 50 mm were placed in front of each detector, and an oscilloscope (DPO71254C, Tektronix, USA) with a sampling rate of 100 GS/s was used to digitize the waveforms. Considering the relatively long decay times of both scintillators, 100,000 data points (corresponding to 1,000 ns in length) per waveform

*Contact author: ryoota@ucdavis.edu, ryosuke.ota@crl.hpk.co.jp

†Contact author: sunkwon@ucdavis.edu

were recorded. The $^{22}$Na point source (~1.18 MBq) was sequentially placed at three different positions.

For the BaF$_2$-IMP experiment, the bandwidth of the oscilloscope was optimized for timing performance and set to 5 GHz. One-layer of polytetrafluoroethylene (PTFE) tape covered the entrance surface to reflect more photons toward the photocathode. Energy deposition was calculated by integrating the waveform over 500 ns from the trigger time. This long integration window was used because some scintillation photons from the slow component were still partially detectable due to the non-zero QE around 300 nm. Energy resolution was analyzed using a function of "a Gaussian + an exponential," which represents a photopeak and a contamination from Compton scattering events, respectively. The resolution was defined as the FWHM (standard deviation × 2.355) of the Gaussian component. An energy threshold of >450 keV was applied to both detectors when evaluating the CTR.

For the timing pick-off algorithm, a spline curve was first applied to the recorded waveforms using the *TSpline3* class method implemented in the ROOT software package [23]. Then, constant-fraction threshold levels ranging from 0.2% to 12.8% of each pulse height were applied in 0.2% increments to correct for time walk effect [24]. For each threshold level, a time difference histogram between the detector pair was generated to determine the optimal threshold level. Each histogram was fitted with a single Gaussian, and the CTR was defined as the FWHM of the Gaussian.

For the BGO-IMP experiment, the bandwidth was set to 3 GHz. In the case of the BGO WFP, black tape was used to cover the entrance surface to suppress temporal fluctuations caused by reflected Cherenkov photons in the BGO plate. Energy deposition was calculated by integrating the waveform over 1,000 ns. The energy resolution, the energy threshold, and the timing pick-off algorithm were defined in the same manner as for the BaF$_2$-IMP. However, a different fitting function was used for the time difference histogram. In this study, we phenomenologically defined the following function to describe the unconventional distribution of the time difference:

$$f(t) = C \left( \sum_{i=0}^{1} \rho_i \frac{e^{-\frac{1}{2}\left(\frac{x-\mu}{\sigma_i}\right)^2}}{\sqrt{2\pi}\sigma_i} + \sum_{i=2}^{3} \rho_i \frac{e^{-\frac{x-\mu}{\tau_i}}}{\tau_i} \vartheta(\mu - x) + \sum_{i=4}^{5} \rho_i \frac{e^{-\frac{x-\mu}{\tau_i}}}{\tau_i} \vartheta(x-\mu) \right) + \text{const}, \quad (1)$$

$$s.t. \; \frac{\rho_2}{\tau_2} + \frac{\rho_3}{\tau_3} = \frac{\rho_4}{\tau_4} + \frac{\rho_5}{\tau_5}, \text{ and } \sum_{i=0}^{5} \rho_i = 1,$$

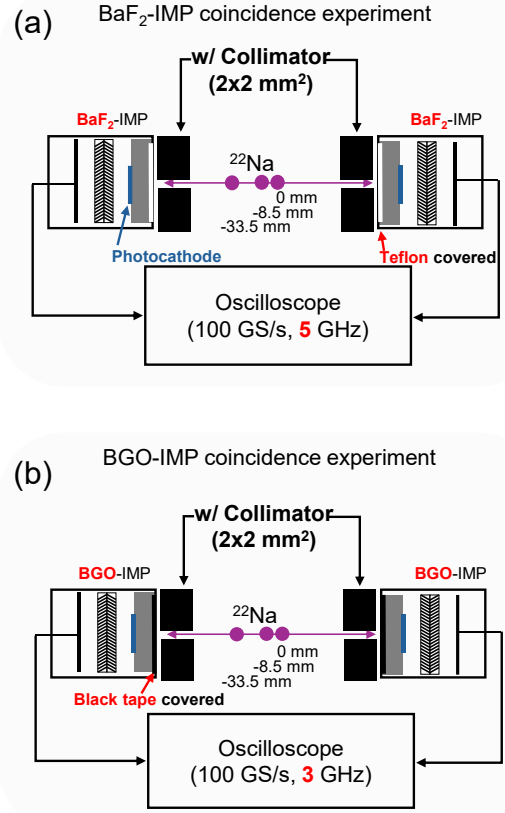

FIG. 2. Experimental setups for CTR evaluation using 511 keV gamma rays produced by a $^{22}$Na point source: (a) a pair of BaF$_2$-IMPs, and (b) a pair of BGO-IMPs. For both experiments, 50 mm thick tungsten collimators were placed in front of each detector. The $^{22}$Na point source was sequentially placed at three different positions. The oscilloscope sampling rate was set to 100 GS/s for both setups, while the bandwidth was optimized depending on the detector type.

where $\mu$ denotes the expected time difference, $\sigma_i$ and $\tau_i$ denote the temporal fluctuations of Cherenkov-based and scintillation-based events, $\rho_i$ denotes their ratio, and $\vartheta(x - \mu)$ denotes the step function, where $\vartheta=0$ if $x < \mu$. Finally, $C$ and const denote coefficients, respectively. The first two Gaussian terms represent events in which both detectors detected at least one Cherenkov photon. The remaining four exponential components represent events in which at least one detector was triggered not by Cherenkov photons but by scintillation photons. The FWHM of the histogram was numerically calculated and defined as the resulting CTR.

*Contact author: ryoota@ucdavis.edu, ryosuke.ota@crl.hpk.co.jp

†Contact author: sunkwon@ucdavis.edu

**C. CTR measurement using cosmic-ray muons**

Investigating the timing performance of the developed detectors using minimum ionizing particles (MIPs) is also of interest because, in many accelerator-based experiments, particles are so energetic (greater than the GeV scale) that they are considered MIPs. Cosmic-ray muons are one of the natural radiation sources and typically have energies exceeding 1 GeV, indicating that CTR evaluation using MIPs is possible even without the use of an accelerator. Given that the peak momentum of cosmic-ray muons from an azimuthal angle ~0° is on the order of 10 GeV/c [25-28], and the energy loss of a charged particle in matter is described by the Bethe–Bloch equation [29], a rough calculation yields an expected energy loss of 3–4 MeV in a 3.2 mm thick $BaF_2$ or BGO crystal.

This significantly larger energy deposition should yield a better CTR compared to that from 511 keV gamma-rays in both materials. In the case of $BaF_2$, this is because CTR is inversely proportional to the square root of light yield (corresponding to energy loss) [13]. In the case of BGO, a greater number of Cherenkov photons is expected, as the number of emitted Cherenkov photons is proportional to the particle's track length in the medium.

For the cosmic-ray muon experiment using a pair of $BaF_2$-IMPs, the detectors were positioned facing each other and placed in close proximity, as shown in Fig. 3(a), considering the relatively low count rate of cosmic-ray muons (~1 count·min$^{-1}$·cm$^{-2}$). In contrast, in the BGO-IMP experiment, only the bottom $BaF_2$-IMP was replaced with the BGO-IMP B (Fig. 3(b)). This is because cosmic-ray muons mostly enter from above and traverse the top detector from its photocathode side. Since Cherenkov photons are emitted along the muon trajectory, they are directed away from the photocathode rather than entering it directly. Therefore, if the top detector is replaced with BGO-IMP, the emitted Cherenkov photons must be reflected at the side opposite to the photocathode before reaching it, which degrades the timing performance. Such a situation does not occur for the $BaF_2$-IMP, which emits scintillation photons isotropically.

The timing pick-off algorithm was the same as that used for the 511 keV gamma-rays. The CTR was calculated from time difference histograms using a fitting function of 'double Gaussians,' which more accurately represented the histogram than a single Gaussian. The energy threshold for the CTR evaluation was set to a 1,500 keV equivalent due to two main difficulties: *(i)* reconstructing the muon's trajectory with only one detector pair and *(ii)* accurately determining the energy deposition from waveforms, given the limited photosensitive area. This corresponds to a charge at least three times greater than that of a 511 keV event. This energy threshold ensured, in an approximate manner, that the cosmic-ray muons passed through the central region of the scintillators.

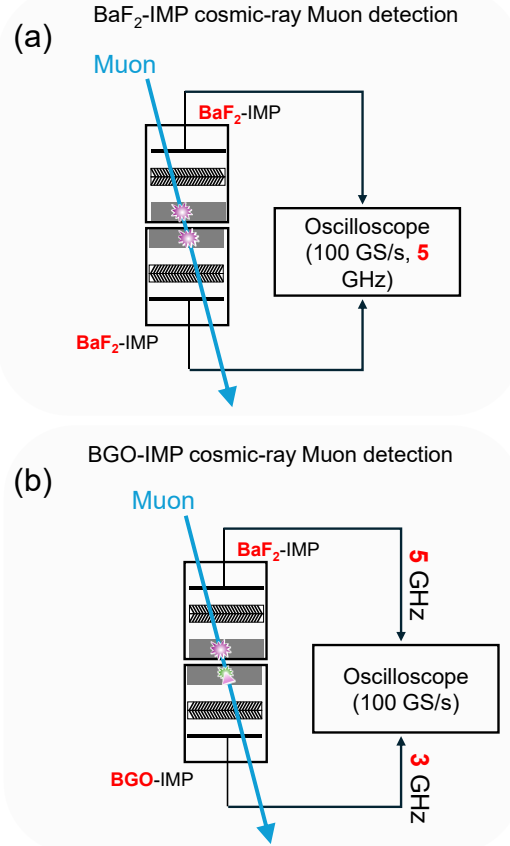

FIG. 3. Experimental setups for CTR evaluation using cosmic-ray muons: (a) a pair of $BaF_2$-IMPs, (b) a BGO-IMP and a $BaF_2$-IMP. The oscilloscope setup was the same as that used in Fig. 2.

### III. RESULTS
**A. Energy resolution at 511 keV**

The energy histograms obtained using 511 keV gamma rays from the $BaF_2$-IMPs are shown in Fig. 4(a), along with those from the BGO-IMPs in Fig. 4(b). Although the slow component of the $BaF_2$ emission was deliberately suppressed, the 511 keV photopeak was clearly observed with corresponding energy resolutions of 35.1 ± 0.6% and 35.2 ± 0.5%. A similar energy resolution of 37.5 ± 0.2% was obtained for BGO-IMP B. However, BGO-IMP A showed a significantly worse energy resolution of 51.2 ± 0.3%. This discrepancy may be attributed to fabrication differences between the two BGO-IMPs.

*Contact author: ryoota@ucdavis.edu, ryosuke.ota@crl.hpk.co.jp

†Contact author: sunkwon@ucdavis.edu

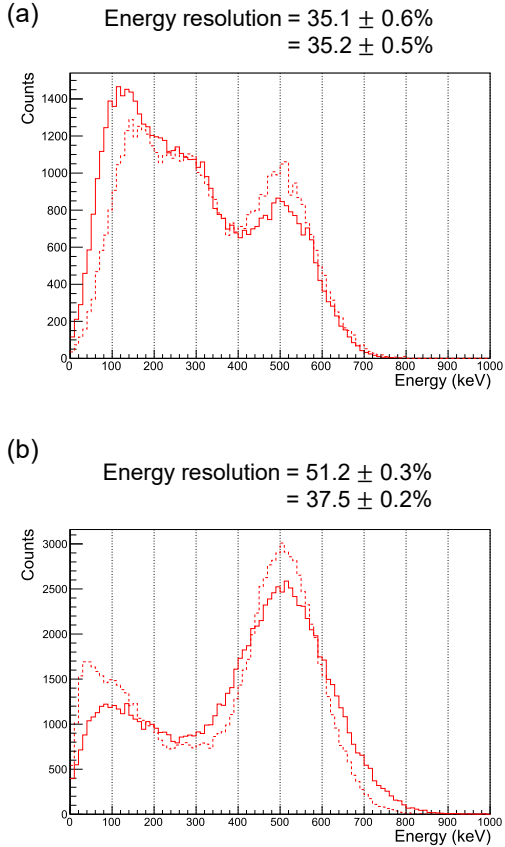

FIG. 4. Energy histograms of (a) BaF$_2$-IMPs and (b) BGO-IMPs measured using 511 keV gamma-rays. For the BaF$_2$-IMPs, both detectors showed the similar energy resolution. In contrast, a significant difference in energy resolution was observed between the two BGO-IMPs, which is attributed to differences in the fabrication process.

### B. CTRs from 511 keV gamma-rays

Figure 5(a) shows the time difference histograms of the BaF$_2$-IMPs for three different source positions. Each histogram was well described and fitted by a single Gaussian, and the peak positions shifted appropriately according to the true source positions. The CTRs from left to right were 52.3 ± 1.2, 54.8 ± 1.3, and 51.5 ± 1.4 ps FWHM, respectively.

The corresponding histograms for the BGO-IMPs are illustrated in Fig. 5(b). Similar shifts in peak positions were observed, as in the case of the BaF$_2$-IMPs. The CTRs from left to right were 46.4 ± 5.5, 48.3 ± 5.1, and 50.1 ± 4.2 ps FWHM, respectively. The full widths at tenth maximum (FWTMs) were calculated as 637.0± 418.6, 984.0± 225.8, and 1229 ± 180 ps, respectively. The large errors were due to the low statistics in the long tail regions.

The entire histogram of the BGO-IMP pair is shown in Fig. 6(a), along with the fitted curve (in red) for the source position at 0.0 mm. As clearly seen, the fitting function (Eq. 1) matches the distribution of the histogram over a 10 ns window. The observed asymmetric distribution, where the left tail is longer than the right, is attributed to differences in the properties between the BGO-IMPs (good vs. poor energy resolution). The $\chi^2/NDF$ value, which is commonly used to assess the goodness of fit, was 0.96. Within the sharp peak region near the center shown in Fig. 6(b), the Cherenkov-based events were clearly represented by double Gaussians. The ratio of Cherenkov photon-triggered events to the total number of coincidence events was calculated to be 8.5 ± 2.1%. This indicates that, in a single detector, approximately 30% of events were triggered by fast Cherenkov photons. The FWHM of the blue solid line in Fig. 6(b), corresponding to the Cherenkov-based event, was 38.2 ps.

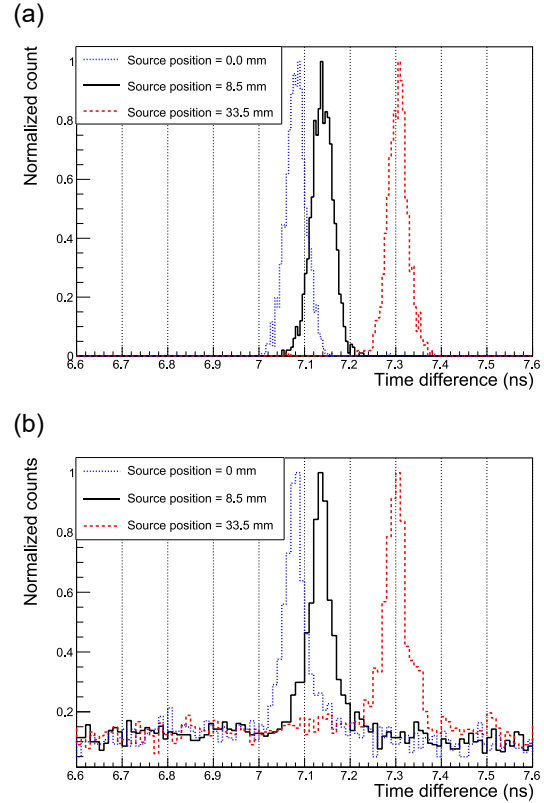

FIG. 5. Histograms of the time difference between detector pairs at three different source positions: (a) BaF$_2$-IMPs and (b) BGO-IMPs. The peak positions shifted appropriately according to the true source positions.

*Contact author: ryoota@ucdavis.edu, ryosuke.ota@crl.hpk.co.jp

†Contact author: sunkwon@ucdavis.edu

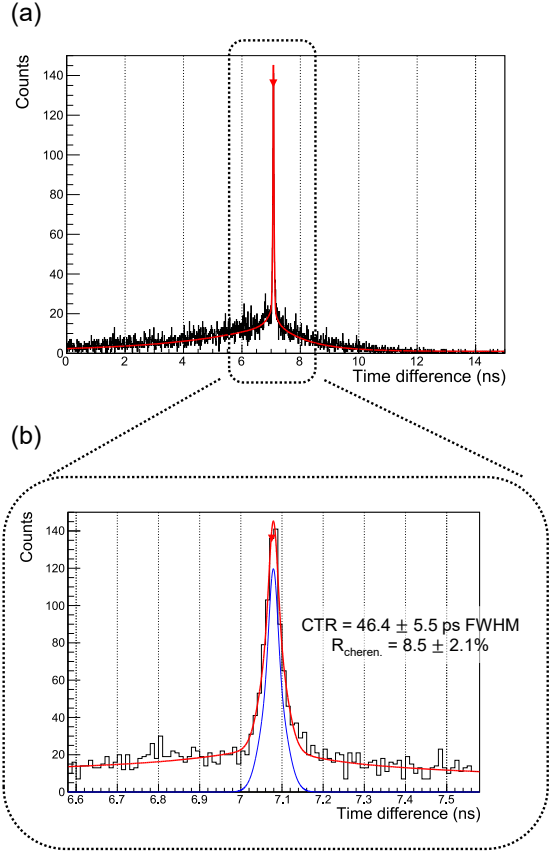

FIG. 6. (a) Full histogram of the time difference between the pair of BGO-IMPs, shown with the red fitting function. The fitting function accurately represents the overall distribution, with a $\chi^2/\mathrm{NDF}$ of 0.96. The central sharp peak, modeled by two Gaussians and labeled as $R_{\mathrm{cheren.}}$ in the figure, accounts for 8.5% of the total area. The FWHM of the blue solid line, corresponding to the Cherenkov-based event, is 38.2 ps.

### C. CTRs from cosmic-ray muons

Cosmic-ray muon-based energy spectra measured with the BaF$_2$- and BGO-IMPs are shown in Fig. 7. Unlike the energy spectra obtained using a $^{22}$Na source, no distinct peaks were observed due to the broad momentum distribution of cosmic-ray muons, and their varying incident angles and positions. The calculated energy loss of 3–4 MeV in the 3.2 mm thickness BaF$_2$ and BGO is consistent with the spectra shown in Fig. 7. Events with energy deposition exceeding 5,000 keV equivalent are attributed to muons entering at relatively large incident angles, resulting in a longer path length in the scintillator crystals.

*Contact author: ryoota@ucdavis.edu, ryosuke.ota@crl.hpk.co.jp

†Contact author: sunkwon@ucdavis.edu

Figure 8(a) depicts the time difference histogram from cosmic-ray muon detection with an energy threshold of 1,500 keV equivalent. The CTR was 25.1 ± 2.8 ps FWHM for the BaF$_2$-IMP pair, showing a significantly better value compared to that obtained with 511 keV gamma rays, due to the greater number of detected photons. By comparison, a CTR of 16.8 ± 4.0 ps FWHM was obtained from the BaF$_2$-IMP and BGO-IMP pair, as shown in Fig. 8(b).

Neither histogram could be described by a single Gaussian, and both exhibited small bumps on the left side of the peak position. These were caused by events in which cosmic-ray muons passed through the central region of the scintillators, resulting in a photon yield exceeding the dynamic range of the oscilloscope. Such saturated events may introduce undesired fluctuations in the measured waveforms [22], thereby producing the bumps observed in the histograms, as shown in the insets of Fig. 9 (around $t$ < 140 ns).

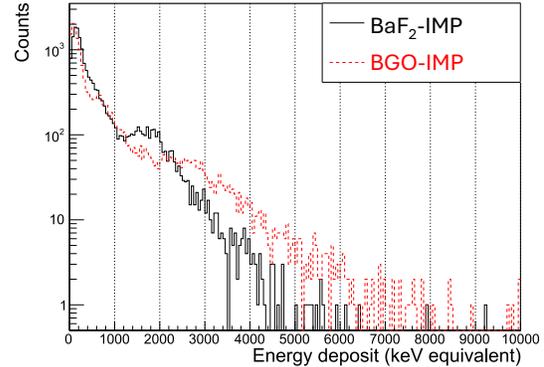

FIG. 7. Energy spectra obtained from the cosmic-ray muon experiment. The energy deposit is broadly distributed up to approximately 10,000 keV equivalent depending on the tracks of cosmic-ray muons. An energy threshold of 1500 keV equivalent was applied for the CTR evaluation.

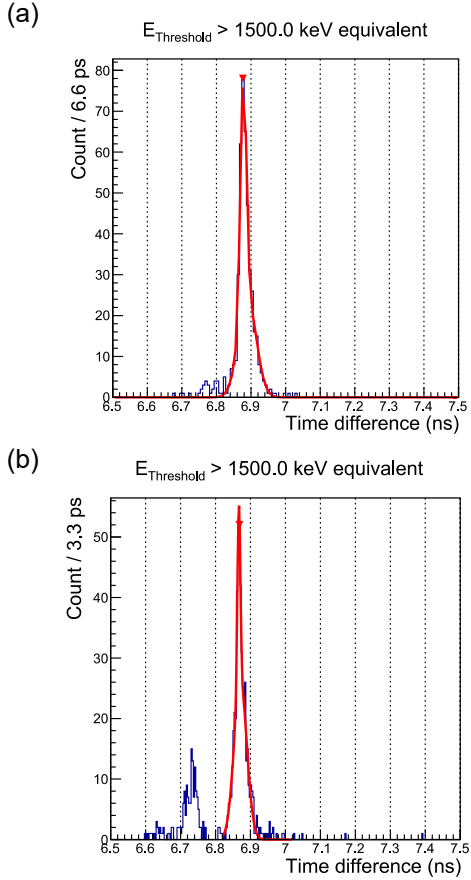

FIG. 8. Time difference histograms with fitting functions based on cosmic-ray muons using (a) a $BaF_2$-IMP pair and (b) a $BaF_2$-IMP and BGO-IMP pair. The energy threshold was set to >1500 keV equivalent, indicating that at least three times more photons were detected compared to the experiments using a $^{22}$Na point source. The CTRs of (a) and (b) were $25.1 \pm 2.8$ and $16.8 \pm 4.0$ ps FWHM, respectively. The bump on the left side of the main peak is attributed to incorrect timing pick-off due to signal fluctuations.

## IV. DISCUSSION

Two types of SCI-IMPs employing $BaF_2$ and BGO scintillators as WFPs were successfully developed for ultrafast timing applications, such as high-energy physics experiments and TOF-PET. Compared to the previously developed CRI-MCP-PMT, SCI-IMPs can simultaneously achieve ultrafast timing performance on the order of a few tens of ps and provide energy measurements. Notably, the timing performance was maintained across a wide energy range, from keV– to GeV–scale particles although the energy deposition of a GeV–scale particle is up to the MeV–scale. Moreover, due to the direct integration of scintillators with photodetectors, there is no need to consider the selection of optimal coupling materials, which is particularly important in the case of $BaF_2$ [15].

The timing performance of the $BaF_2$ scintillator was fully exploited through the integration technique. The $BaF_2$-IMP can push its timing performance further by increasing the number of detected photons, as the CTR theoretically improves with the inverse square root of the number of detected photons [13]. The most limiting factor in the timing capability of the $BaF_2$-IMP is its limited photosensitive area. A simple solid angle calculation subtended by the photosensitive area from the center of the crystal shows that the light collection efficiency is ~35% under the current configuration. Therefore, further enhancement of the CTR can be expected by increasing the photosensitive area.

$BaF_2$ may not be a practical choice for future TOF-PET systems due to its relatively low stopping power for 511 keV gamma rays [30]. However, it remains under consideration for use in high-energy physics experiments [4], and this study has demonstrated a CTR as low as 25 ps FWHM. Furthermore, the suppression of the slow components in $BaF_2$ by the CsTe photocathode enables a high-count rate capability, which is advantageous for high-energy physics experiments.

Regarding the timing performance of the BGO-IMP pair, to the best of our knowledge, this is the first experimental result demonstrating a BGO-based CTR below 50 ps FWHM using 511 keV gamma rays, which is more than three times faster than previously reported in other studies using similar thicknesses of BGO [13, 31]. This improvement can be attributed to the optimal propagation of Cherenkov photons to the photocathode and the superior SPTR of MCP-PMTs compared to silicon photomultipliers. As the probability of Cherenkov photon detection increases with higher energy deposition in BGO, the CTR can be further improved, reaching down to $16.8 \pm 4.0$ ps FWHM for the cosmic ray muon experiment. Consequently, BGO can be considered a fast hybrid Cherenkov/scintillator material, despite having been largely set aside since the emergence of lutetium-based scintillators.

This study also provides an experimental observation regarding the emission characteristics of BGO. Most research on the CTR of BGO-based detectors uses a double-Gaussian function to represent the time difference histogram (e.g., [13, 32-34]), because either a Cherenkov or a scintillation photon can trigger a photodetector as the first photon, causing the time difference histogram to deviate from a single Gaussian. However, the time difference histogram obtained using the BGO-IMP pair, as shown in Fig. 6(a), cannot be explained using multiple Gaussians due to its exponential-like tails. These tails arise from the nature of the probability density function (PDF) of individual scintillation photons from BGO. Considering the theoretical PDF of BGO emission based on [35], the shape of the PDF for the first emitted photon is asymmetric (**APPENDIX**). In addition, the fast SPTR does not smear the shape of the PDF, indicating that the histogram explicitly reflects the shape of the PDF. This interpretation is further supported by a Monte Carlo simulation performed in this study (**APPENDIX**), as the simulated time difference histogram also exhibited a similar distribution that can be represented by a combination of Gaussians and exponentials. However, the reason why four exponential components, rather than two, are required to fit the histogram in Fig. 6(a) remains under investigation.

*Contact author: ryoota@ucdavis.edu, ryosuke.ota@crl.hpk.co.jp

†Contact author: sunkwon@ucdavis.edu

Regarding the detection efficiency of BGO-IMPs for 511 keV gamma-rays, the stopping power of BGO is higher than that of both $BaF_2$ or the lead glass used in CRI-MCP-PMTs. Therefore, higher detection efficiency can be achieved. However, considering the ratio of Cherenkov-Cherenkov events, which form the sharp peak in Fig. 6(a), relative to the total number of events, the choice of the most suitable detector remains debatable. Methods that can efficiently extract Cherenkov events should be developed to leverage the Cherenkov-only CTR of 38.2 ps FWHM and maximize the potential of BGO-IMPs [31, 33, 38]. This limitation is less relevant in cosmic-ray muon experiments due to the higher probability of detecting Cherenkov photons, which eliminates the nanosecond-order long tail.

Figure 9(a, b) shows the averaged waveforms of the BGO-IMP B for the experiments using 511 keV gamma rays and cosmic-ray muons, from which the average numbers of detected Cherenkov and scintillation photons can be quantified. For the 511 keV gamma ray experiment, the average numbers of detected Cherenkov and scintillation photons are calculated to be 0.862 and 134 photons, respectively. This result indicates that Cherenkov photons are not always detected by the BGO-IMP pair. Consequently, scintillation-triggered events account for 91.5% of the time difference histogram as depicted in Fig. 6(a).

However, by applying a threshold on the initial photon density [31] of 0.085 pC in the first 200 ps, the ratio of the fast component, denoted as $R_{Cheren}$, was increased from 8.5% to 20%, thereby reducing tails caused by scintillation-triggered events. Furthermore, Fig. 10(a) shows that the CTR improves with the photon density threshold, achieving 28.4 ps FWHM at 0.095 pC. The CTR of 28.4 ps outperforms the Cherenkov-based CTR of 38.2 ps FWHM shown in Fig 6(b). This implies that more than one Cherenkov photon is detected at the photon density threshold of 0.095 pC on average, thereby improving the effective SPTR. It is worth noting that applying the photon density threshold improves the FWTM more quickly than the FWHM. This observation indicates that scintillation-based slow events are efficiently classified at a low photon density threshold, which preserves the remaining Cherenkov-based fast events up to 0.03 pC (Fig. 9(b)). Thus, the photon density threshold is a powerful tool for event classification.

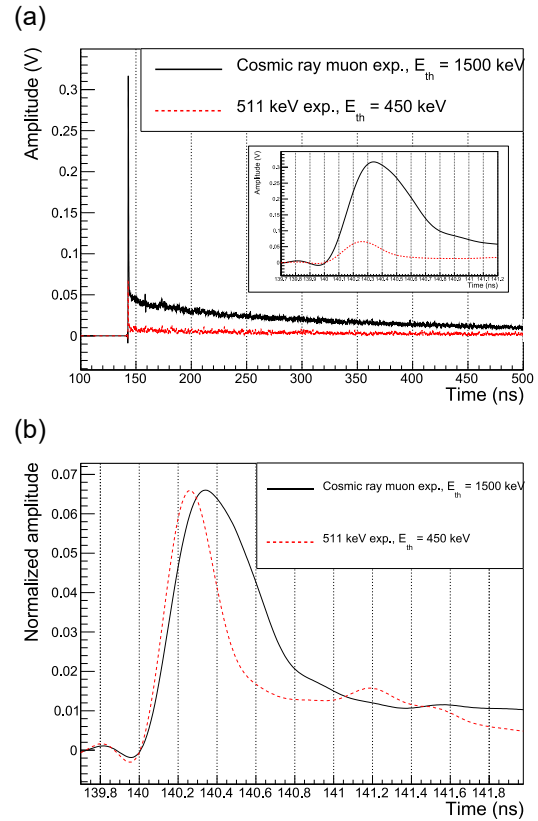

FIG. 9. (a) Overall averaged waveforms for the 511 keV gamma ray and cosmic-ray muon experiments. The prompt rising part from 140 to 141 ns, caused by Cherenkov photon detection, can be clearly seen. (b) Normalized average waveforms zooming in on the rising part. Interestingly, the temporal promptness of Cherenkov photon detection for the 511 keV experiment appears faster than that for the cosmic-ray muon experiment.

*Contact author: ryoota@ucdavis.edu, ryosuke.ota@crl.hpk.co.jp

†Contact author: sunkwon@ucdavis.edu

In contrast, for the cosmic-ray muon experiment, the average numbers of detected Cherenkov and scintillation photons are 10.5 and 762 photons, respectively. The significantly higher number of detected Cherenkov photons in this case results in improved CTR compared to that obtained using 511 keV gamma rays. Moreover, an average of 10.5 Cherenkov photons is sufficient to suppress scintillation-triggered events, thereby eliminating the long tails in the time difference histogram (Fig. 8(b)). Intriguingly, the temporal promptness of the Cherenkov-induced rising part is noticeably faster in the case of 511 keV experiment than in the case of the cosmic-ray muon experiment when zooming in on the rising parts of the averaged waveforms in Fig. 9 (b). This can be attributed to the following two factors: *(i)* rapid energy deposition of photoelectrons in the BGO WFP until their energy falls below the Cherenkov threshold, and *(ii)* variability in the incident angles of cosmic-ray muons combined with the large (60 degrees) opening angle of the Cherenkov cone, which results in a wide range of Cherenkov photon paths in the BGO WFP.

The current form factor of the SCI-IMPs has two limitations: a relatively small photosensitive area compared to the scintillator dimensions and the bulkiness (large dead space surrounding the sensitive area) of the current single-channel MCP-PMTs. In particular, the bulkiness is problematic for both high-energy physics experiments and TOF-PET applications, where it limits detector packing fraction. However, there is a clear way to address both limitations simultaneously. Developing a two-dimensional, position-sensitive, multi-anode SCI-IMP would drastically increase the photosensitive area, allow denser detector alignment, and make the detector arrays more compact while simultaneously achieving better spatial resolution. Although this fabrication technology is highly challenging and still in its early stages, recent work by another group [39] has demonstrated its feasibility using a pure Cherenkov radiator. In addition, the QE of the SCI-IMP can be technically improved by 30% or more through optimization of the photocathode type, which would in turn enhance the energy and timing resolutions as well as increase the ratio of Cherenkov-based fast events. Building on this, our ongoing work involves further development and characterization of multi-anode SCI-IMPs with ultrahigh timing resolution, aiming to make them more practical.

To further improve the timing performance of SCI-IMPs, faster scintillators such as photonic-crystal, composite, and perovskite scintillators are promising candidates for future ultrafast timing applications [40-42]. By integrating available and emerging technologies, we anticipate the realization of direct positron emission imaging (dPEI) systems as a possible ultimate form of PET [43-45].

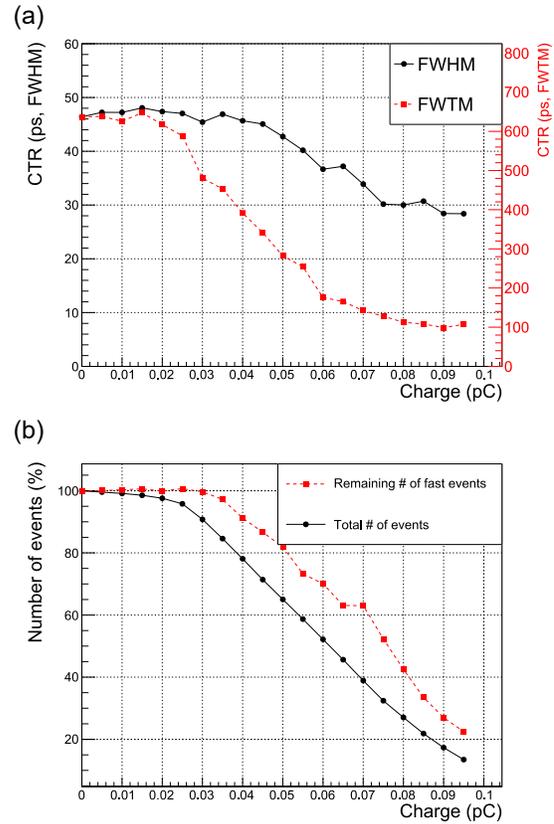

FIG. 10. (a) CTRs of FWHM and FWTM as a function of the photon density threshold defined as charge detected within the first 200 ps. Applying the photon density threshold first improves the FWTM, and then improves FWHM. (b) Relative number of total and fast (Cherenkov-based) events as a function of the photon density threshold in coincidence. Applying the photon density threshold up to 0.03 pC can preserve the number of fast events while efficiently classifying the slow (scintillation-based) events.

### V. CONCLUSIONS

We developed SCI-IMPs using BGO and BaF$_2$ scintillators that simultaneously achieve fast timing and energy measurement, indispensable for high energy physics experiments and TOF-PET. For performance evaluation, 511 keV electron-positron annihilation gamma rays and cosmic-ray muons were used to cover the energy range from keV to GeV. For the 511 keV experiment, the energy resolutions were measured to be 35.1 ± 0.1% and 35.2 ± 0.2% for the two BaF$_2$-IMPs, and 37.5 ± 0.2% and 51.2 ± 0.3% for the two BGO-IMPs. The CTRs were 52.3 ± 1.2 ps for the BaF$_2$-IMP pair and 46.4 ± 5.5 ps for the BGO-IMP pair. In comparison, for the cosmic-ray muon experiment, the CTRs were obtained as 25.1 ± 2.8 and 16.8 ± 4.0 ps, respectively, when the energy threshold was set to >1500 keV. The developed SCI-IMPs achieved timing performance on the order of a few tens of ps while being capable of energy measurement. Through the developed versatile integration technique and further optimizations, SCI-IMP is expected to broaden its applicability and, in particular, serve as essential detectors for future high energy physics experiments and medical imaging.

*Contact author: ryoota@ucdavis.edu, ryosuke.ota@crl.hpk.co.jp

†Contact author: sunkwon@ucdavis.edu

**ACKNOWLEDGMENTS**
This work was supported by the National Institutes of Health grant R01 EB033536. The authors appreciate the colleagues in the Electron Tube Division, Hamamatsu Photonics K.K., for their contribution to the assembly of the BaF$_2$- and BGO-IMPs.

**APPENDIX: Theoretical calculation of the PDF of BGO and Monte Carlo simulation**

For the theoretical calculation of the PDF of BGO based on [35], the emission kinetics of BGO, excluding Cherenkov emission, were modeled with two decay time constants:

$$f(t) = \sum_{i=1}^{2} R_i \frac{e^{-t/\tau_{\text{decay\_i}}} - e^{-t/\tau_{\text{rise}}}}{\tau_{\text{decay\_i}} - \tau_{\text{rise}}},$$

where $R_i$ and $\tau_{\text{decay\_i}}$ denote the scintillation light yield and the decay time constants of the fast and slow scintillation components, respectively. $\tau_{\text{rise}}$ denotes the common rise time constant. The values of $\tau_{\text{decay\_1}}, \tau_{\text{decay\_2}}, \tau_{rise}$, and the ratio of the fast component to the total light yield were set to 41.8, 282.5, 0.00569 ns, and 6.62%, respectively. These numbers were obtained by performing a time correlated single photon counting method, where a BGO crystal with dimensions of 21.85 mm$\phi$ × 3.2 mmt was used.

For simplicity, the SPTR of an MCP-PMT is assumed to form a single Gaussian distribution, and the temporal response $r(t)$ can be expressed as follows:

$$r(t) = \frac{1}{\sqrt{2\pi}\sigma} e^{-\frac{1}{2}\left(\frac{t-\bar{t}}{\sigma}\right)^2},$$

where $\bar{t}$ denotes the expected time from photon detection to signal readout, and $\sigma$ denotes the SPTR. Therefore, the temporal distribution $f'(t)$, which includes both scintillation emission and photodetector detection, can be described by:

$$f'(t) = (f * r)(t),$$

where * denotes the convolution operator. The SPTR for the SCI-IMPs used in this study was assumed to be 22 ps FWHM based on [22]. The probability $P_N(t)$ that $N$ photons are emitted within time $t$ is given by:

$$P_N(t) = \frac{\left(\int_0^t f'(t')dt'\right)^N e^{-\int_0^t f'(t')dt'}}{N!}.$$

Consequently, the probability $W_Q(t)$ that the $Q^{\text{th}}$ photon is emitted within the interval ($t$, $t + dt$) is expressed as:

$$W_Q(t)dt = P_{Q-1}(t) \times f'(t)dt$$

Fig. 11(a) illustrates the PDF of the first emitted photon in BGO, showing a distinctly asymmetric distribution. Furthermore, Fig. 11(b) depicts the simulated histogram of the time difference between the first-photon arrival times of a pair of BGO-IMPs modeled in the Geant4. The red dotted and blue dashed lines represent double-Gaussian and Gaussian-plus-exponential fits, respectively. It is clearly observed that the double-Gaussian function cannot accurately model the obtained distribution in the region around ±0.3 ns. The $\chi^2$/NDF values for the double-Gaussian and Gaussian-plus-exponential fits are 1.27 and 0.94, respectively.

*Contact author: ryoota@ucdavis.edu, ryosuke.ota@crl.hpk.co.jp
†Contact author: sunkwon@ucdavis.edu

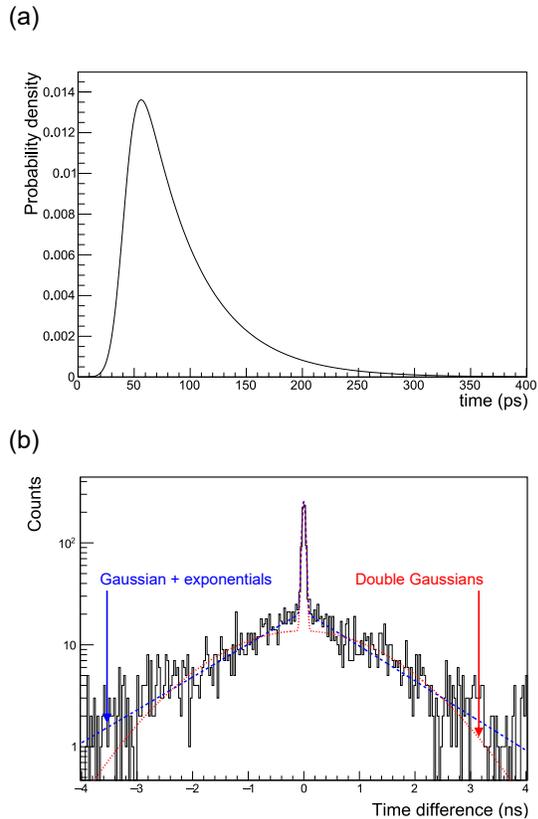

FIG. 11. (a) Theoretically calculated PDF of the first emitted photon in BGO, convolved with a Gaussian-shaped SPTR of 22 ps FWHM. The resulting PDF exhibits a clearly asymmetric distribution. (b) Simulated histogram of the time difference between a pair of BGO-IMPs. The red dotted and blue dashed lines represent double-Gaussian and Gaussian-plus-exponential fits, respectively. The latter provides a more accurate representation of the histogram.

*Contact author: ryoota@ucdavis.edu, ryosuke.ota@crl.hpk.co.jp

†Contact author: sunkwon@ucdavis.edu

*Contact author: ryoota@ucdavis.edu, ryosuke.ota@crl.hpk.co.jp

†Contact author: sunkwon@ucdavis.edu

*Contact author: ryoota@ucdavis.edu, ryosuke.ota@crl.hpk.co.jp

†Contact author: sunkwon@ucdavis.edu